\documentclass[preprint]{rsl}

\title{Unlocking the Hidden Potential of ALMA\\
Calibrators: A Pilot Blind Search for Galactic\\
Molecular Gas in ALMA Band 1
}
\author{Kanako Narita and Bunyo Hatsukade}

\begin{document}

\maketitle
\begingroup
\renewcommand\thefootnote{}\footnotetext{\footnotesize Manuscript received 26 March 2026.
This work was supported by the Japan Science and Technology Agency (JST) SPRING
(grant JPMJSP2108), by JSPS KAKENHI Grant-in-Aid for Scientific Research
(grant JP23K03449), and by JSPS Grant-in-Aid for JSPS Fellows (grant JP26KJ0864),
and by the ANRI Fellowship. K. Narita was also supported by the International
Graduate Program for Excellence in Earth-Space Science at the University of Tokyo.

K. Narita is with the Department of Astronomy, Graduate School of Science,
University of Tokyo, 7-3-1 Hongo, Bunkyo-ku, Tokyo 113-0033, Japan; e-mail:
naritakn@g.ecc.u-tokyo.ac.jp.

B. Hatsukade is with the National Astronomical Observatory of Japan, 2-21-1
Osawa, Mitaka, Tokyo 181-8588, Japan; also with the Graduate Institute for
Advanced Studies, 2-21-1 Osawa, Mitaka, Tokyo 181-8588, Japan; and also with
the Department of Astronomy, Graduate School of Science, University of Tokyo,
7-3-1 Hongo, Bunkyo-ku, Tokyo 113-0033, Japan.}
\endgroup

%
%

\begin{abstract}
  We present a pilot blind absorption line search using calibrator observations with the Atacama Large Millimeter/submillimeter Array (ALMA) Band 1 receiver (35 GHz to 50 GHz). Radio-loud quasars, commonly used for calibration, provide stable and bright background sources that are ideal for absorption line studies of diffuse interstellar gas. To enable a systematic analysis beyond previous targeted studies, we developed an automated pipeline, from archival data download to statistical identification of absorption line candidates, and applied it to archival ALMA Band 1 observations corresponding to 16 distinct calibrator sources. We statistically identify significant absorption line candidates along the analyzed sightlines. This pilot study demonstrates the feasibility of blind absorption line searches with ALMA Band 1 calibrator data and provides a scalable framework for future large-scale archival surveys.
\end{abstract}

\section{Introduction}

Galactic molecular absorption lines provide a powerful probe of diffuse interstellar gas that is difficult to detect in emission, particularly the so-called carbon monoxide-dark molecular gas \cite{Lucas1995}. Absorption line studies have commonly exploited radio-loud quasars used as calibrators, which provide bright continuum background sources. Using such calibrators, previous surveys at millimeter wavelengths, including those with the Atacama Large Millimeter/submillimeter Array (ALMA), have shown that diffuse molecular gas is widely distributed along essentially random galactic sightlines \cite{Ando2016,Narita2024, Klitsch2019, Kanekar2014}. However, absorption line searches using ALMA calibrator observations have, so far, primarily targeted redshifted molecular absorption associated with background systems. Although galactic molecular absorption has occasionally been reported as a by-product of these studies [4], dedicated blind searches explicitly optimized for galactic absorption along calibrator sightlines remain limited. Blind millimeter-wave absorption surveys have also been conducted with single-dish telescopes such as the Green Bank Telescope at frequencies overlapping with ALMA Band 1 \cite{Huang2016}, but these surveys have mainly focused on redshifted molecular absorption, whereas the narrow, synthesized beam of an interferometer enables higher effective sensitivity to weak absorption features toward compact background sources, even without requiring spatially resolved information. The newly available ALMA Band 1 covers low-frequency millimeter wavelengths at which rotational transitions of complex, polyatomic molecules, potentially related to the chemical origins of life, are expected \cite{DiFrancesco2013}. This frequency range, therefore, offers a promising opportunity to extend absorption line studies toward diffuse and translucent galactic gas. To enable a systematic exploration of Band 1 calibrator data, we developed an automated analysis pipeline for blind absorption line searches. In this article, we present a pilot blind absorption line survey using ALMA Band 1 calibrator observations. As a first step toward a large-scale Band 1 absorption line survey, our aim is to demonstrate the feasibility of automated blind searches and to assess the scientific potential of Band 1 for probing diffuse interstellar chemistry.

\section{Observation and Analysis Method}
\subsection{ALMA Band 1 Calibrator Data}
We use ALMA Band 1 calibrator observations obtained from the public archive as a first step toward a blind absorption line survey. The analysis is restricted to data from Cycle 11, which provide relatively homogeneous observing setups and manageable data volumes suitable for uniform processing. Roughly 70\% of ALMA calibrators are radio-loud quasars, offering bright and compact continuum emission that is ideal for detecting foreground absorption features along galactic sightlines. In this pilot study, we analyze 20 Band 1 data sets corresponding to 16 distinct calibrator sources. The sample was selected to ensure sufficient continuum signal-to-noise ratio and continuous spectral coverage within the frequency range of Band 1, enabling reliable absorption line searches without prior knowledge of line frequencies or source properties.

\subsection{Automated Analysis Pipeline}
An automated analysis pipeline was developed to enable a systematic and uniform blind search for absorption lines in ALMA Band 1 calibrator data. The scientific design of the pipeline, including spectral extraction, continuum normalization, and detection criteria, was originally developed by the author on the basis of single-sightline analyses and subsequently generalized and automated. The implementation was carried out in a collaborative framework, building upon existing almaqso analysis utilities \cite{Nishida2026}. As illustrated in Figure 1, the researcher configures the analysis settings, including target sources selected from the ALMA calibrator catalog \cite{ALMACalCatalog} (https://almascience.nao.ac.jp/alma-data/calibrator-catalogue), observing band and imaging settings, in a concise script of approximately 10 lines using the almaqso framework. Publicly available data sets matching the specified settings are automatically searched and retrieved from the ALMA archive \cite{ALMAArchive} (https://almascience.nao.ac.jp/aq/) using astroquery \cite{Ginsburg2019}. For each data set found, the pipeline sequentially performs downloading, calibration, imaging, and export as Flexible Image Transport System files using the Common Astronomy Software Applications (CASA) package \cite{CASA2022}. Calibration is performed without masking absorption features of the calibrators, in contrast to the standard calibration scripts distributed by the observatory, which mask such features; this approach preserves genuine spectral signatures for absorption line analysis. Imaging parameters such as pixel scale and spectral windows are automatically determined on the basis of observational parameters, including angular resolution and spectral coverage. The pipeline also includes automated error handling on the basis of CASA log inspection, whereby failed processing outputs are automatically removed and warning messages are issued to prevent incomplete products from remaining in the workflow. Successfully processed data sets are recorded to enable resumable processing and efficient restart after interruptions. One-dimensional quasar spectra are subsequently extracted from the resulting spectral data cubes.

\subsection{Blind Search Strategy}
Prior to the blind search, continuum normalization is performed using a dedicated script developed by the author. Absorption-free spectral regions are automatically identified via sigma clipping, and the mean flux density measured in these regions is used to normalize the spectrum, yielding a continuum-normalized spectrum suitable for statistical line detection. The blind search is conducted without assuming specific molecular species, rest frequencies, or line widths. Candidate spectral features are identified on the basis of statistical significance in continuum-normalized spectra. Both emission and absorption features are symmetrically identified, with a peak signal-to-noise ratio of 5$\sigma$ or higher adopted as the primary detection criterion. Adjacent significant channels are grouped to define emission or absorption line candidates. For each candidate, an integrated signal-to-noise ratio over the grouped channels is evaluated, and only features with an integrated signal-to-noise ratio of 5$\sigma$ or higher are retained as validated detections. The integrated signal-to-noise ratio is computed assuming a constant channel width within each spectrum. Features exhibiting strong emission-like or absorption-like signatures that appear consistently across multiple, unrelated sightlines are attributed to artificial signals or instrumental artifacts and are therefore excluded from the analysis. The remaining detections are identified solely on the basis of statistical significance and are treated as candidate features rather than confirmed astrophysical lines. Spectral channels at the bandpass edges, where the sensitivity is significantly reduced, are excluded to minimize spurious detections. In this pilot study, the emphasis is placed on demonstrating the feasibility and robustness of blind absorption line searches in ALMA Band 1, rather than on detailed molecular identification or physical modeling.
\begin{figure}
  \centering
  \includegraphics{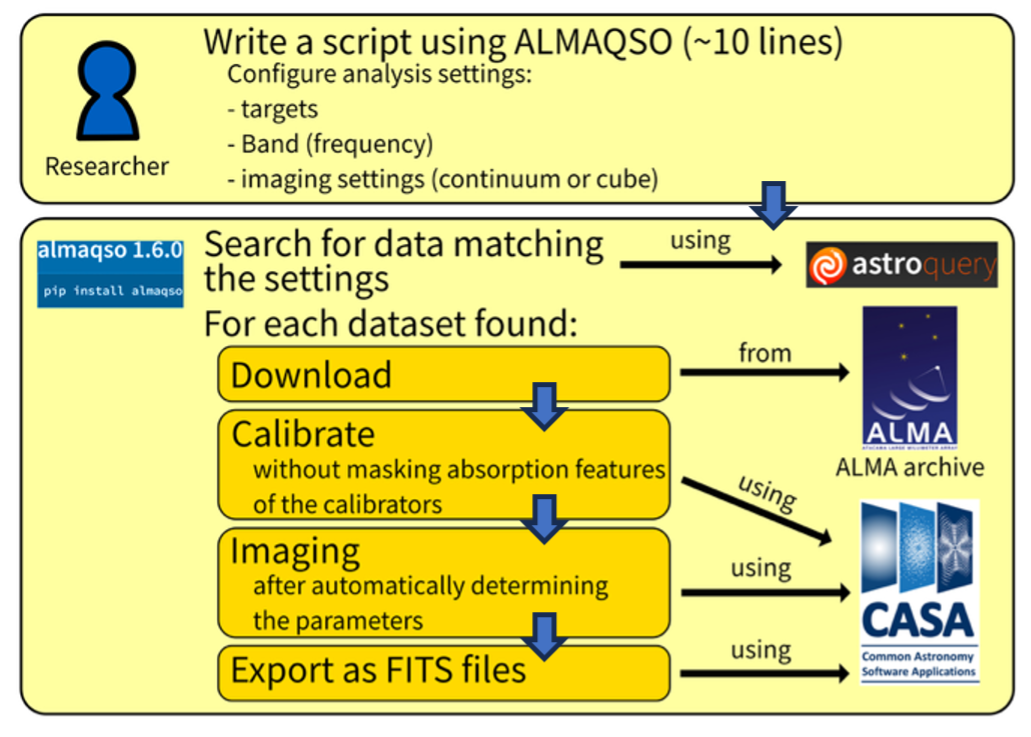}
  \caption{Overview of the automated analysis pipeline used in this study. Public ALMA calibrator data sets are retrieved from the ALMA archive using astroquery, calibrated and imaged with CASA, and processed through the almaqso framework to extract one-dimensional spectra and perform a blind absorption line search.}
  \label{fig:worlflow}
\end{figure}

\section{Results}
After applying the blind search criteria and quality control procedures, the automated pipeline identifies a single robust line candidate across 16 analyzed sightlines. The detected candidate is found toward J1744-3116, whose line of sight passes close to the galactic plane and the inner Galaxy, where the column density of diffuse molecular gas is expected to be significantly higher than along higher galactic latitude sightlines. In contrast, the remaining calibrators probe predominantly high-latitude directions with substantially lower molecular gas columns, naturally reducing the probability of detectable absorption. We present the J1744-3116 sightline exhibiting the most significant detection in our survey. Figure 2 shows the continuum-normalized spectrum toward J1744-3116, where an absorption-like feature is identified near 48.99 GHz. The absorption feature is detected at a significance of approximately 9$\sigma$ and extends over several consecutive spectral channels. The measured line depth and width are consistent with absorption arising from diffuse galactic interstellar gas \cite{Gerin2021}. In the present pilot survey, the automated framework processed archival ALMA observations corresponding to 16 distinct sightlines from 6 observing projects in approximately 13 h with minimal user intervention, substantially reducing the need for repetitive interactive operations and manual error checking required in conventional analysis workflows.

\begin{figure}
  \centering
  \includegraphics[width=\linewidth]{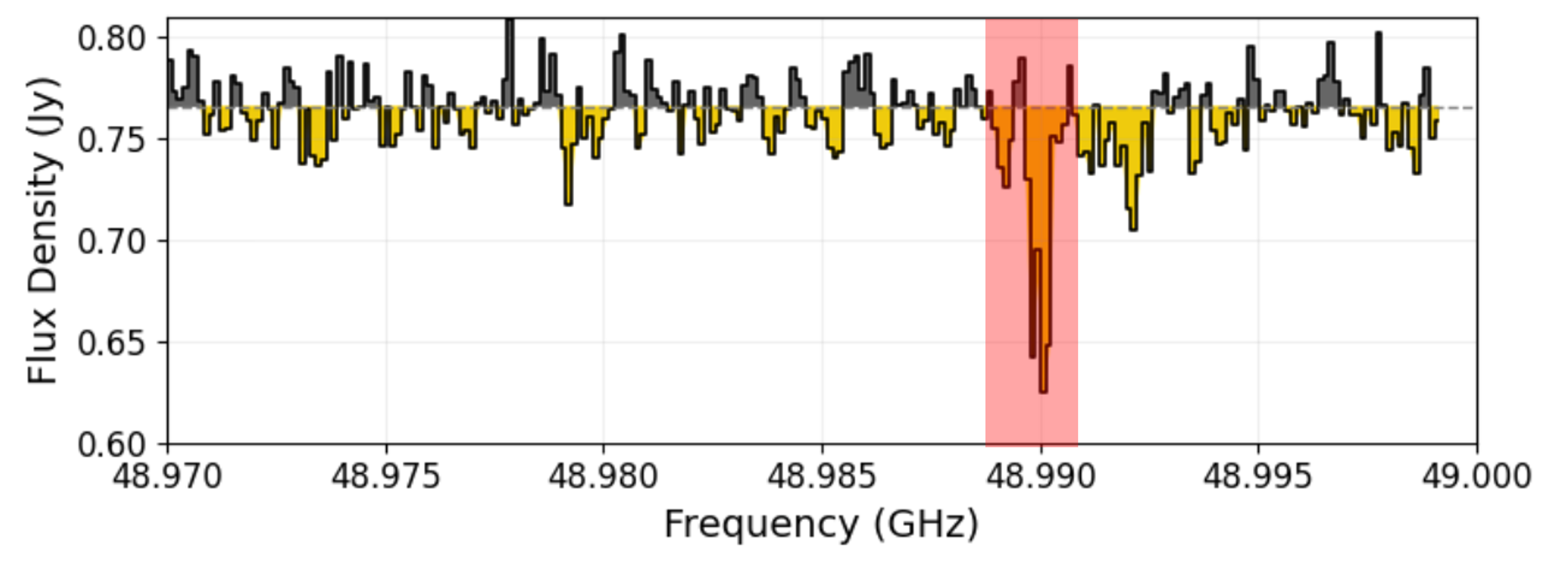}
  \caption{Example spectrum of a calibrator sightline toward J1744-3116 observed in ALMA Band 1, showing the most significant absorption-like feature identified by the blind search. The absorption feature spans multiple adjacent spectral channels near 48.99 GHz.}
  \label{fig:spe}
\end{figure}

\section{Discussion}
\subsection{Implications for ALMA Band 1 Absorption Studies}
This pilot study demonstrates that ALMA Band 1 calibrator observations can be effectively used for blind absorption line searches toward diffuse galactic gas. The detection of statistically significant absorption line candidates confirms that Band 1 provides sufficient sensitivity and spectral quality to probe poorly explored low-frequency molecular transitions. Combined with the availability of numerous bright quasar calibrators, Band 1 offers a promising new window for systematic absorption line studies of diffuse and translucent interstellar chemistry.
\subsection{Role of Automated Analysis and Future Prospects}
The use of an automated analysis pipeline enables uniform and reproducible processing of large calibrator data sets, overcoming the limitations of previous targeted or manual studies. Although this pilot survey is restricted to a subset of data from Cycle 11, the same approach is readily scalable to future Band 1 observations with larger data volumes and more diverse observing configurations. The current analysis is limited by sample size and by uncertainties in molecular identification. Future work will expand the data set to additional observing cycles and incorporate follow-up analyses to identify molecular species and derive physical properties of the absorbing gas.

\subsection{Conclusions}
We conducted a pilot blind search for galactic molecular absorption lines using ALMA Band 1 calibrator observations from Cycle 11. By developing and applying an automated and reproducible analysis pipeline, we demonstrated that Band 1 calibrator data can be systematically and uniformly analyzed without prior assumptions on molecular species, rest frequencies, or line widths. After applying stringent blind search criteria designed to ensure robustness against noise fluctuations and instrumental effects, a single robust absorption line candidate was identified. Although limited in the number of detections, this result demonstrates the feasibility of blind absorption line surveys in frequency ranges that have not been extensively explored. This study establishes ALMA Band 1 calibrator observations as a promising resource for probing diffuse interstellar gas and complex molecular chemistry, and the automated approach presented here provides a scalable framework for future large-scale absorption line surveys using ALMA calibrator archives.

\section{Acknowledgments}
K.N. acknowledges Akimasa Nishida and Ryo Kishikawa for valuable scientific discussions and the ALMA project and its staff for enabling the observations used in this study.
This work was supported by the Japan Science and Technology Agency (JST) SPRING (grant JPMJSP2108), by JSPS KAKENHI Grant-in-Aid for Scientific Research (grant JP23K03449), by JSPS Grant-in-Aid for JSPS Fellows (grant JP26KJ0864), and by the ANRI Fellowship.
K.N. was also supported by the International Graduate Program for Excellence in Earth-Space Science at the University of Tokyo.
This paper makes use of the following ALMA data: ADS/JAO.ALMA\#2024.1.00211.S, ADS/JAO.ALMA\#2024.1.00412.S, ADS/JAO.ALMA\#2024.1.00671.S, ADS/JAO.ALMA\#2024.1.00950.S, ADS/JAO.ALMA\#2024.1.01045.S, ADS/JAO.ALMA\#2024.1.01125.S and ADS/JAO.ALMA\#2011.0.00001.CAL.
ALMA is a partnership of ESO (representing its member states), NSF (USA) and NINS (Japan), together with NRC (Canada), NSTC and ASIAA (Taiwan), and KASI (Republic of Korea), in cooperation with the Republic of Chile. The Joint ALMA Observatory is operated by ESO, AUI/NRAO and NAOJ.


\begin{thebibliography}{99}

\bibitem{Lucas1995}
R. Lucas and H. S. Liszt, ``The Physics and Chemistry of Interstellar Molecular Clouds,''
in G. Winnewisser and G. C. Pelz (eds.), \emph{The Physics and Chemistry of Interstellar Molecular Clouds},
Berlin, Springer, 1995, p. 120.

\bibitem{Ando2016}
R. Ando, K. Kohno, Y. Tamura, T. Izumi, H. Umehata, et al.,
``New Detections of Galactic Molecular Absorption Systems Toward ALMA Calibrator Sources,''
\emph{Publications of the Astronomical Society of Japan}, \textbf{68}, no. 1, February 2016, p. 6.

\bibitem{Narita2024}
K. Narita, S. Sakamoto, J. Koda, Y. Yoshimura, and K. Kohno,
``Physical and Chemical Properties of Galactic Molecular Gas Toward QSO J1851+0035,''
\emph{The Astrophysical Journal}, \textbf{969}, no. 2, July 2024, p. 102.

\bibitem{Klitsch2019}
K. Klitsch, C. P\'eroux, M. A. Zwaan, I Smail, D. Nelson, et al.,
``ALMACAL---VI. Molecular Gas Mass Density Across Cosmic Time via a Blind Search for Intervening Molecular Absorbers,''
\emph{Monthly Notices of the Royal Astronomical Society}, \textbf{490}, no. 1, November 2019, pp. 1220--1230.

\bibitem{Kanekar2014}
N. Kanekar, A. Gupta, C. L. Carilli, J. T. Staocke, and K. W. Willett,
``A Blind Green Bank Telescope Millimeter-Wave Survey for Redshifted Molecular Absorption,''
\emph{The Astrophysical Journal}, \textbf{782}, no. 1, February 2014, p. 56.

\bibitem{Huang2016}
Y. D. Huang, O. Morata, P. M. Koch, C. Kemper, Y. J. Hwang, et al.,
``The Atacama Large Millimeter/Submillimeter Array Band 1 Receiver,''
\emph{Modeling, Systems Engineering, and Project Management for Astronomy VII},
Edinburgh, UK, June 26--July 1, 2016, p. 99111V.

\bibitem{DiFrancesco2013}
J. Di Francesco, D. Johnstone, B. Matthews, N. Bartel, L. Bronfman, et al.,
``The Science Cases for Building a Band 1 Receiver Suite for ALMA,''
arXiv preprint arXiv:1310.1604 [astro-ph.IM], 2013. https://doi.org/10.48550/arXiv.1310.1604 (Accessed May 27, 2026).

\bibitem{Nishida2026}
A. Nishida, K. Ryo, Y. Yoshimura, and K. Narita,
\texttt{almaqso}, version 1.6.0,
Geneva, Switzerland, Zenodo, 2026, doi:10.5281/zenodo.18334952.

\bibitem{ALMACalCatalog}
Joint ALMA Observatory,
``ALMA Calibrator Catalogue,''
\texttt{https://almascience.nao.ac.jp/alma-data/}\\
\texttt{calibrator-catalogue} (Accessed May 27, 2026).

\bibitem{ALMAArchive}
Joint ALMA Observatory,
``ALMA Science Archive,''
\texttt{https://almascience.nao.ac.jp/aq/} (Accessed May 27, 2026).

\bibitem{Ginsburg2019}
A. Ginsburg, B. Sipo{\dag}cz, C. E. Brasseur, P. Cowperthwaite, P. S. Craig, et al.,
``astroquery: An Astronomical Web-Querying Package in Python,''
\emph{Astronomical Journal}, \textbf{157}, no. 3, March 2019, p. 98.

\bibitem{CASA2022}
CASA Team, B. Bean, S. Bhatnagar, C. Sandra, D. Meyer, et al.,
``CASA: Common Astronomy Software Applications,''
\emph{Publications of the Astronomical Society of the Pacific}, \textbf{134}, no. 1041, November 2022, p. 114501.

\bibitem{Gerin2021}
M. Gerin and H. Liszt,
``CO$^+$ as a Probe of the Origin of CO in Diffuse Interstellar Clouds,''
\emph{Astronomy \& Astrophysics}, \textbf{648}, April 2021, p. A38.

\end{thebibliography}
\end{document}